\documentclass[conference, a4paper]{IEEEtran}
\usepackage{graphicx}
\usepackage{amsmath,amssymb,epsfig,xcolor} % easier math formulae, align, subequation
\usepackage{subfig}
\usepackage{tabularx,multirow,array}
\usepackage{times}
\usepackage{booktabs}
\usepackage{multirow}
\usepackage{soul}
\usepackage{enumerate}
\usepackage{xcolor}
\usepackage[countmax]{subfloat}
\usepackage{algpseudocode}
\usepackage{algorithm}
\usepackage{threeparttable}
\usepackage{cite}
\usepackage{url}
\usepackage{amsthm}
\usepackage{bbm}
\theoremstyle{plain}

\SetMathAlphabet{\mathcal}{bold}{OMS}{cmsy}{b}{n}
\begin{document}

\title{Benchmarking Various ML Solutions in Complex Intent-Based Network Management Systems\\
%{\footnotesize \textsuperscript{*}Note: Sub-titles are not captured in Xplore and
%should not be used}
%\thanks{Identify applicable funding agency here. If none, delete this.}
}

\author{\IEEEauthorblockN{Mounir Bensalem, Jasenka Dizdarevi{\'{c}}, and Admela Jukan}
\IEEEauthorblockA{Technische Universit\"at Braunschweig, Germany \\
%\textit{name of organization (of Aff.)}\\
%City, Country \\
\{mounir.bensalem, j.dizdarevic, a.jukan\}@tu-bs.de}

}

\maketitle

\begin{abstract}
Intent-based networking (IBN) solutions to managing complex ICT systems have become one of the key enablers of intelligent and autonomous network management.  As the number of machine learning (ML) techniques deployed in IBN increases, it becomes increasingly important to understand their expected performance. Whereas IBN concepts are generally specific to the use case envisioned, the underlying platforms are generally heterogenous, comprised of complex processing units, including CPU/GPU, CPU/FPGA and CPU/TPU combinations, which needs to be considered when running the ML techniques chosen. We focus on a case study of IBNs in the so-called ICT supply chain systems, where multiple ICT artifacts are integrated in one system based on heterogeneous hardware platforms. Here, we are interested in the problem of benchmarking the computational performance of ML technique defined by the intents. Our benchmarking method is based on collaborative filtering techniques, relying on ML-based methods like Singular Value Decomposition and Stochastic Gradient Descent, assuming initial lack of explicit knowledge about the expected number of operations, framework, or the device processing characteristics. We show that it is possible to engineer a practical IBN system with various ML techniques with an accurate estimated performance based on data from a few benchmarks only.
\end{abstract}

%\begin{IEEEkeywords}
%...
%\end{IEEEkeywords}

\section{Introduction}\label{intro}

Managing, orchestrating and controlling environments which include distributed devices spanning highly heterogeneous networks is becoming an increasingly complex task. For this reason, the interest in more intelligent and automation geared management methods, such as utilization of intent based network (IBN) management, has also been on the rise, and is significantly improving both the automation and management process in wide spectrum of application domains and complex ICT systems.  IBN systems are designed to translate high-level intents into a structured format with the help of natural language processing techniques, thus easing the system management also for non-experts. Structured specifications translated from the intent configure the systems in a policy-based fashion,  without the need to manually check the conflicts in a large-scale system. IBN typically deploy various ML techniques not only for natural language processing, but also for the entire suite of network management functions, such as resource management or security.

%There has been an increasing demand for adopting machine learning based solutions in the myriad of complex ICT systems, as they are expected to significantly improve both the automation and management process in wide spectrum of application domains. These applications are often assumed to be time critical, such as prompt analysis of security threats, incidents and suspicious activities. Furthermore, with the hardware developments in the edge and IoT domain of novel ICT systems, which are now capable of running highly resource demanding ML algorithms, the heterogeneity of the hardware has increased. This has particularly been the case in ICT supply chain application domain, with its highly diverse  infrastructure, requirements and capabilities. To deploy and test existing and new ML models in said environments,  dependence of software performance on the host hardware has to be taken into consideration, with new models and hardware being continually added. 

%ighly beneficial to actually estimate in advance the execution performance of the ML solution to be used.
%Complication

Whereas IBN concepts are generally specific to the use case envisioned, the underlying ICT platforms are generally heterogenous, comprised of complex processing units, including CPU/GPU, CPU/FPGA and CPU/TPU combinations. Hence, for the system to choose between multiple potentially new ML model configurations is not only potentially time-critical, but can also result in computationally ineffective choice of ML techniques on a specific hardware. Moreover, intents are useful precisely because they enable system configurations where a problem can be solved by choosing from a set of candidate ML-based solutions, whereby each of the candidates can give the comparable output, but with different accuracy and run-time. Hence, an efficient IBN system needs to be able to predict how long it takes for a potential ML solution to execute for a given intent, while preserving a threshold of accuracy. This brings into question of what the optimal way of estimating performance of ML solutions in an IBN system is, and whether it is possible to learn and extract features that relate the performance of a specific intent to the hardware in an implicit way.  What is needed is an intent-based networking paradigm that would allow administrators to choose and deploy best ML models on the most suited hardware platform, through simple use of high level intents.

%Solution
The objective of this paper is three-fold. Firstly, we introduce a specific case study of complex IBN management systems implemented in ICT supply chain systems. The implementation of intents in a specific case study enables us to define meaningful intents, which in turn allows for an easier system administration. Second, we focus on the design of the ML benchmarking solution for intents, using a specific module within IBN-based management system, we refer to as \emph{ML Recommender}. This module estimates the computational performance of ML-based techniques on heterogeneous hardware devices in the system, without explicit knowledge about the number of operations, framework, or the device processing characteristics of each ML technique. To discover and learn the features that link the performance of a specific ML model to the hardware in an implicit way we will use Collaborative Filtering (CF), as a commonly used tool in recommendation systems \cite{delimitrou2013paragon}. Finally, we evaluate the system performance by measuring the quality of forecasting using a small number of tests on some devices, and  the information from previous execution of other ML techniques on existing hardwares. The information collected about the performance is represented as  a sparse matrix. The matrix would contain information about the computational performance of certain ML tasks, such as Deep Learning models, which were  hosted on specific hardware devices in previous executions. After that, ML algorithm called Singular Value Decomposition (SVD), will be used in order to estimate the missing values of the input matrix, which will in turn be used as valid ML-performance estimation. We show how the estimated outcome effectively enhances IBN-based management systems capabilities of scheduling, planning or hardware provisioning. 

The rest of this paper is organized as follows. Section \ref{sec:1} describes the related work. Section  \ref{sec:2} describes the case study of intent-based management in ICT supply chains. The ML benchmarking solution is presented in \ref{sec:MLrec}. Section \ref{sec:evaluation} evaluates the performance. We conclude the paper in Section \ref{sec:conclusion}.

\section{Related work} \label{sec:1}

With a rise of complex network environments and system architectures, there has also been an increased interest in more intelligent and autonomous management methods, with intent based networking as one of the leading trends.  Intent based networking is a practical concept used in specific reference scenarios. In \cite{su12072782} and \cite{8004109}, multi-platform 5G and IoT network infrastructures are studies with IBN solutions. In \cite{9200928} distributed, multi-technology, and multi-stakeholder network infrastructures are taken as examples to drive the need for intelligent automation and orchestration. Our previous work in \cite{bensalem2021role} introduced and motivated intent-based networking in the so-called ICT supply chain system, defined as \emph{a system integrating ICT products and services, transforming raw materials,
and components into a finished product or service from supplier
to the end user.}. Similarly, a part of closed-loop automation was analyzed in supply chains, with different management  system requirements formulated with intents \cite{gomes2021intent}. 

Every IBN architecture deploys ML techniques, which in general need to run on rich heterogeneous hardware platform environments.  Hence, for an intent to be executed effectively, it is critical to understand the performance of the ML solution deployed. In \cite{nemirovsky2017machine}, an ML approach based on artificial neural network was proposed, with the aim of predicting the performance of different applications and with that advancing the effectiveness of scheduling on a CPU hardware. In \cite{makrani2019xppe}, the execution time of an application on a specific FPGA is used to predict the execution time on other hardware without making a real implementation, using neural networks as a tool for estimation. In addition, software developers need to define some system characteristics of their developed components before deploying in production. In  \cite{hafnaoui2016simulation}, a methodology is presented that estimates the execution time of software components on a specific architecture using simulation and analytical tools that uses parts of system information. The paper compared the estimated execution times of certain software components using the simulation based system and the real benchmarks. The work presented in \cite{delimitrou2013paragon} proposes an online and scalable datacenter scheduler, considering the  heterogeneity of hardware and the interference between executed workloads, using the collaborative filtering prediction technique. The same technique will be used in this paper, due to its wide use in recommendation systems. One of the most famous applications of CF is Netflix Challenge \cite{bell2008bellkor}, which will also be used as a reference method in our work, as well as the framework developed in \cite{delimitrou2013paragon}. 

\section{A Case Study of Intent-Based Management in ICT Supply Chains} \label{sec:2}

As ICT supply chain networks are comprised of heterogeneous and complex infrastructures, with the high interoperability requirements. Here, managing, orchestrating and enforcing policies at scale is a challenge. To tackle this challenge and develop an intelligent management strategy, we employ the paradigm of intent-based networking in order to allow administrators to set high level intents that instruct the system of \emph{what to do} instead of \emph{how} a task should be executed.

\subsection{IBN Reference Architecture}
\begin{figure*}
  \centering
    \includegraphics[scale=0.43]{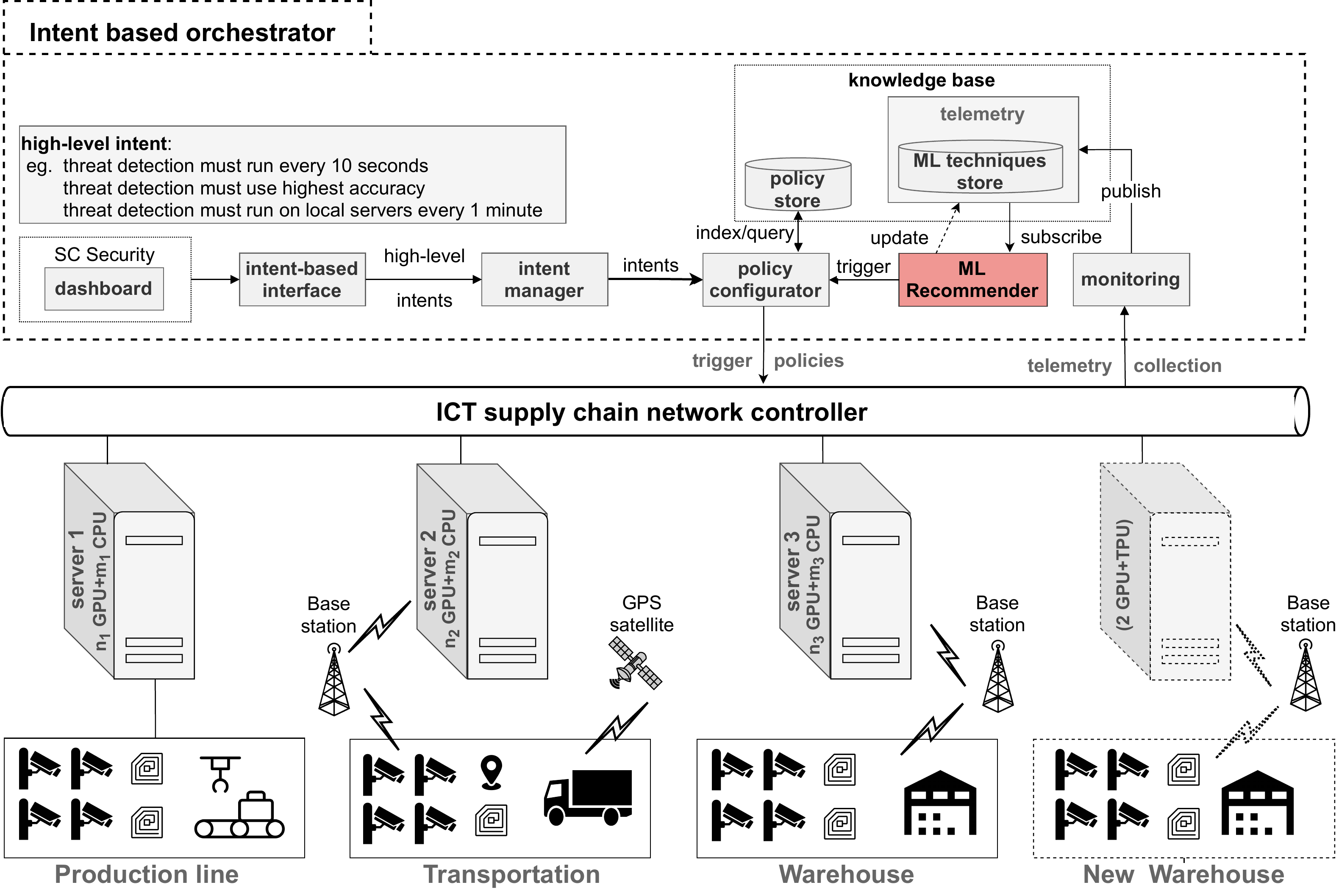}
     %\vspace{-0 cm}
  \caption{Reference architecture of an intelligent intent based management system in ICT supply chain}
 \label{fig:sysarch}
 \vspace{-0.2 cm}
 \end{figure*}

The case study of IBN based management system design is illustrated in Fig.~ \ref{fig:sysarch}, with the ML-based intelligence as the key enabler of the proposed solution, considering the need for automation of system configurations. The ICT supply chain network typically consists of several domains, such as factories, transportation facilities, warehouses, and retailers. Each domain can have its own computing servers, which are used for various computational functionalities, such as networking, security, system optimization, and system management. Tasks can be forwarded to dedicated servers depending on the required computation, memory, throughput etc. 
%One of the main knowledge discovery approaches that our system can use will be designed, explained and implemented in this paper.

The reference architectural design of the intelligent intent based management system is illustrated in Fig.~\ref{fig:sysarch}. The system consists of ICT supply chain network infrastructure, including various IoT devices such as security cameras, RFIDs, and sensors, edge computing nodes located geographically close to the IoT devices,  a centralized ICT supply chain network controller to monitor and manage the infrastructure, and a high-level intent-based orchestrator. The network controller collects telemetry from the infrastructure, manage and enforce policies in the system. The intent-based orchestrator is a high-level  layer that interacts with the network administrator in order to provide an easy-to-use management interface.  It includes different modules that insures the translation of user, i.e., system administrator, intents into configured policies to be applied by the network controller: a dashboard, intent-based interface, intent manager, policy configurator, knowledge base, monitoring, and ML recommender. The \textit{dashboard}   provides a user interface, where the intents given by the network administrator are read and then parsed by an \textit{intent-based interface}. An \textit{intent manager} component is designed to translate high-level intents into a structured format with the help of natural language processing techniques. Structured specifications translated from the intent are then forwarded to the  \textit{policy configurator} component, which matches the user requirements with possible predefined policies stored in a \textit{knowledge base}, and more specifically in a \textit{policy store}. After configuring these policies, conflicts are checked and verified. Then the \textit{policy configurator} triggers the network controller to apply changes in the network infrastructure. Monitoring tools are used by the network controller to collect metrics and provide the real time state of the network. The \textit{Monitoring} component in the intent-based orchestrator filters the telemetry collected from the network and uses the \textit{knowledge base} to store  some selected telemetry such as CPU/GPU performance, memory, latency, throughput, security state of links and devices, application logs, and security camera results etc. One important telemetry information is the the ML techniques performance on different hardware, which is saved in an \textit{ML techniques store}. In order to use this telemetry and improve policy configuration methods such as scheduling of functions on servers, routing, and enhancing access control configuration, we design an intelligent module called \textit{ML Recommender}. This module will tackle one of the most important challenges related to choosing and deploying the best ML-based algorithms in the most suitable server of the supply chain infrastructure, using the performance information collected in the \textit{ML techniques store}.\\

\subsection{Benchmarking ML Solutions in IBN Systems}\label{sec:ML-in-ibn}
ML solutions are widely used for various tasks in ICT supply chains, such as threat detection, demand forecasting, and face detection.
 When a new device joins the network, it has no performance history, which makes the system enable to accurately take several decisions such as ML techniques allocation. To solve this problem,  the  \textit{ML Recommender} is introduced, which is responsible on benchmarking the performance of ML techniques on new coming device or the performance of new ML techniques on existing devices. (The next section describes the design of the ML recommender in detail.) From a system design side, in order to trigger the  \textit{ML Recommender}, we propose two intent types that can be used by the system administrator, as follows: 
\begin{itemize}
\item For adding a new hardware device: "add device $\textit{device\_id}$ to domain  $\textit{domain\_id}$"
\item For adding a new ML technique: "add ML-technique $\textit{ML-tech\_id}$ to ML-technique type  $\textit{ML-tech\_type\_name}$"
\end{itemize}
The intent manager then discovers the intent types from the input text and parse it in a structured JSON format as: \{intent\_name : "adding device", device: "device\_id", domain: "domain\_id"\}, and \{intent\_name : "adding ML-technique", ML-technique: "ML-tech\_id", ML-technique\_type: "ML-tech\_type\_name"\}. After that the policy configurator matches the given intent with the required policies. For instance, this can be defined as follow: check the connectivity of device ("device\_id"), or check the existence of ML-technique ("ML-tech\_id") in an ML techniques store using its "ML-tech\_type\_name", alert the user if the checking operation fails, trigger the \textit{ML Recommender} with the provided information, or similar.\\

  \section{ML Benchmarking}\label{sec:MLrec}

As a part of the intelligent intent based management system, \textit{ML Recommender} module will have the task of benchmarking and estimating the performance of different ML techniques on heterogeneous ICT supply chain hardware platforms.  Instead of running all existing ML techniques on each device, to link the performances of a specific ML model to the hardware we will use Collaborative Filtering.  This section first explains the CF techniques envisioned,  after which we present how they are being applied to estimate technique-device benchmarks.

\subsection{Collaborative Filtering Techniques}\label{subsec:1-1}

We adopt a single rating CF system for product recommendation, where the input is modeled as a sparse matrix $E\in \mathbb{R}^{m,n}$, where $m$ is the number of users and $n$ is the number of items. Each user $i$ is represented by  one row and each item $j$ by one column. In this example, we denote by $e_{i,j}$ the rating of item $j$ by user $i$. A new incoming user will  rate a limited number of items, then the existing user rating is used to estimate the ratings for other missing items.  CF algorithms ensure that the predicted value is always in the same range used by all users. After obtaining the complete rating information, a recommendation system recommends the list of Top-N most likable items by the user.

One of the efficient methods of CF, which among many other recommender systems was also used in the Netflix Challenge \cite{bell2008bellkor}, is the Singular Value Decomposition SVD, that we will adopt as a first step of our estimation approach. SVD is a technique of dimensionality reduction used to decompose a matrix into three matrices $U$, $V$, and $\Sigma$, in order to find lower bi-dimensional feature space.

\begin{equation}\label{eq:svd}
SVD(E)= \begin{pmatrix}
 e_{1,1}& .  &.  & .&e_{1,n}\\ 
 e_{2,1}& . &. &. &e_{2,n} \\ 
 .&  . &.  &. &. \\
 .& . & e_{i,j} & . &. \\ 
 .&  . &.  &. & .\\
e_{m,1}& . & . & . &e_{m,n}
\end{pmatrix} =U\Sigma V^{T}
\end{equation}

Where $U$ and $V$ are $m \times r$and  $n\times r$ orthogonal matrices, representing the left and right singular vectors respectively, and $\Sigma$ is an $r \times r$ diagonal matrix, representing the singular values, $r$ denotes the rank of matrix $E$, and expresses the number of features of similarity resulted by SVD.
The application of SVD requires the factorization of our input matrix $E$, which is a complex problem due the sparsity of the matrix. Traditional SVD algorithms cannot work with incomplete information about the entries and in addition they cause the issue of overfitting, meaning that the estimation will converge to some known values from the given input. \\ In order to recover missing entries of the matrix $E$, PQ-reconstruction method \cite{delimitrou2013paragon} will be used. To that end we define the matrices $Q=U$, $P^T=\Sigma V^{T}$, and $R=Q\times P^T$ as the approximation of A, including the missing entries. The vector associated to the item $j$ is denoted by $q_j\in \mathbb{R}^r$, which expresses the features representation of item $j$.  Similarly for users,  $p_i\in \mathbb{R}^r$ denotes the vector associated to the user $i$, to express its features representation. Therefore, the elements of the approximation matrix $R$ can be represented as:
\begin{equation}\label{eq:r}
r_{i,j}=q_j^T p_i
\end{equation}
The estimation of any rating of a user $i$ to an item $j$ can be easily obtained, once we have the feature vectors $p_i$ and $q_j$.

In matrix reconstruction, the estimation model is built using the observed ratings and avoiding the missing entries, represented by zero values in the sparse matrix $E$. 
For the overfitting problem, a regularization term is used  while minimizing the error during the learning phase. The following minimization formulation of the problem tries to minimize the squared error between the rating and the estimation, considering the regularization term:
 \begin{equation}\label{eq:min}
 \min_{q,p} \sum_{(i,j)\in S} (r_{i,j} - q_j^T p_i)^2 + \lambda (||q_j ||^2+ (||p_i||^2)
 \end{equation}
Where $S$ represents the set of non-zero values of matrix $E$, and $\lambda$ is a regularization parameter. The non-zero values are considered to model the previously observed ratings in order to create a general fitting function able to predict unknown ratings, while minimizing the error. The regularization parameter $\lambda$ is used to adjust the minimization function and prevent overfitting the model with known values.

Next, we adopt the Stochastic Gradient Descent as a learning algorithm that processes the matrix $E$ to improve the estimation. The feature vectors $p_i$ and $q_j$ are initialized randomly and iterated over all the training set of ratings. In each iteration the value of the rating using eq. (\ref{eq:r}) is predicted and computed as follows:
\begin{equation}\label{eq:epsilon}
\epsilon_{i,j}=r_{i,j}-q_j^T p_i
\end{equation}
Afterwards, we update  the feature vectors $q_j$ and $p_i$ using a learning rate $\sigma$, considering the regularization factor $\lambda$, as follows: \\
\begin{equation}\label{eq:q}
q_j = q_j+\sigma \cdot (\epsilon_{i,j} \cdot p_i - \lambda \cdot q_j)
\end{equation}
\begin{equation}\label{eq:p}
p_i = p_i+\sigma \cdot (\epsilon_{i,j} \cdot q_j - \lambda \cdot p_i)
\end{equation}
The algorithm iterates over all matrix entries s times, where s is a parameter to be defined, as well as the learning rate $\sigma$, regularization factor $\lambda$, and the number of features $r$ (called latent factors).  
The utilization of the previously described algorithm to estimate the performance of a ML-based technique is described in the next subsection.
%\hl{this subsection would benefit from some exampels that lead back to the idea of ML benchmarking. Is hard to connect it to ML benchmarking...mayeb a numerical example?}
%%%%%%%%%%
\subsection{Predicting Performance of Various ML Techniques}\label{sub-sec:similarity}
\begin{figure}
 \centering
   \includegraphics[scale=0.55]{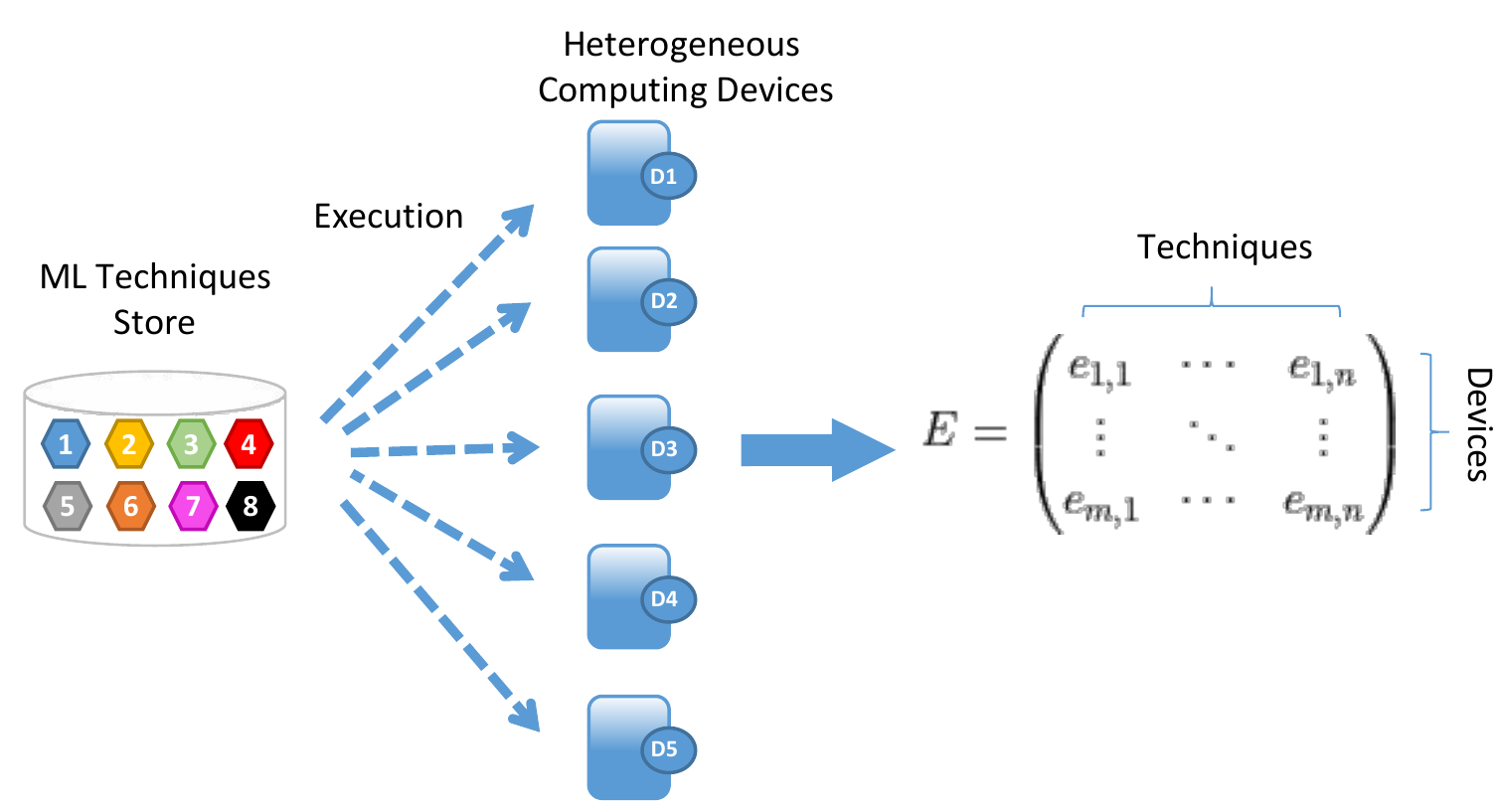}
 \caption{Running ML technique on heterogeneous computing devices}
\label{fig:idea}
\vspace{-0.5cm}
\end{figure}

In proposed intelligent intent based management system, we assume the existence of $M$  ML techniques, that can be executed on $N$  hardware devices. Example of ML-based techniques can be various object detection algorithms used to monitor security cameras, executing different tasks like employee detection, threat detection, product quality monitoring, etc. Different ML models can do the same task, while having different speed performance in inference, memory consumption, and accuracy of results. One of the most important factors that need to be measured is the inference performance, which will be the main objective of our proposed solution to predict. Fig. \ref{fig:idea} represents an abstraction of our problem, where we assume the existence of an ML techniques store containing ML models and several heterogeneous computing devices. The previously explained matrix 
$E$ of user-item ratings in this system represents the execution performance of matching pairs of devices and ML technique pairs. When a new ML technique is registered throught the intent-based orchestrator, as described in section \ref{sec:ML-in-ibn} e.g. \textbf{"add ML-technique $\textit{MobileNet-V2-threat\_1}$ to ML-technique type  $\textit{threat-detection}$"}, we add a new raw to the performance matrix $E$ with empty zeros. After that, the \textit{ML Recommender} chooses randomly a set of devices for the benchmarking of new ML technique.  The collected performance information is then added to $E$ as shown in Fig. \ref{fig:estimation}. a., then the SVD method is used to estimate the performance on the rest of devices using eq.\ref{eq:svd}, \ref{eq:p}, \ref{eq:q}, and \ref{eq:epsilon} in order to minimize the error defined in eq. \ref{eq:min}. Finally the \textit{ML Recommender} saves the new performance entries into the  \textit{ML techniques store}.   \\

\subsection{System Scaling}
\begin{figure}
 \centering
   \includegraphics[scale=0.55]{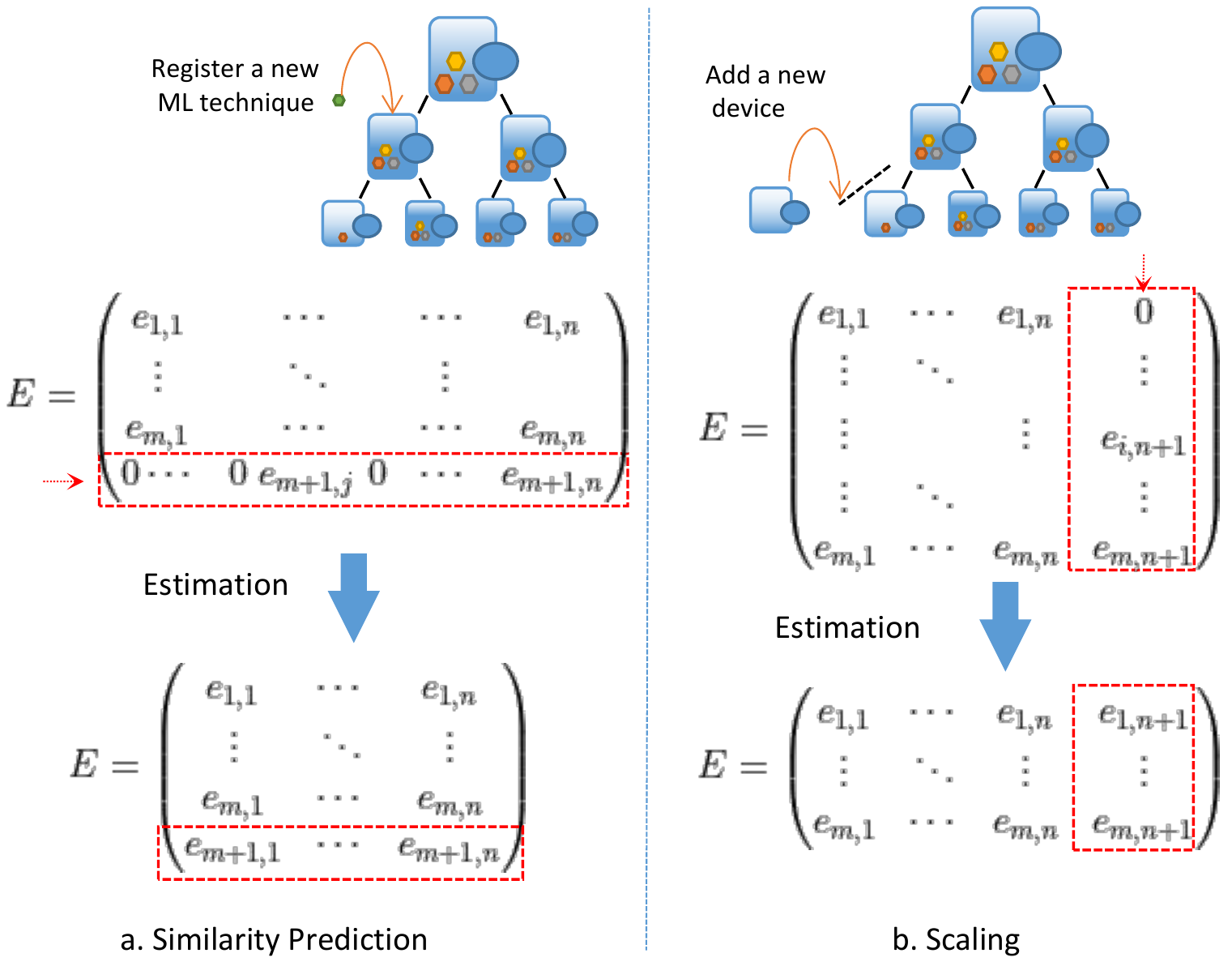}
 \caption{Performance estimation of a ML techniques on a newly added device.}
\label{fig:estimation}
\vspace{-0.2 cm}
\end{figure}

An important factor to be considered in an IBN system design, which also is its salient feature, is the scalability. To this end, we consider a scenario where a new hardware device is added to the system. We start with the assumption that there is no any prior knowledge about the performance of ML techniques on a newly added device. The procedure of performance estimation in this case, is similar to the procedure used for a new incoming technique, triggered by the user from the intent-based orchestrator, e.g. \textbf{"add device $\textit{edge\_100}$ to domain  $\textit{warehouse\_5}$"}. The estimation process again follows two phases, warm-up and online, where the warm-up phase is the same as used previously in Fig. \ref{fig:idea}. When a new device is added, the performance of existing ML-based techniques on that device needs to be benchmarked.  In Fig.\ref{fig:estimation}.b, an illustration of this process is shown. Assuming that $k$ techniques are chosen from the ML techniques store,  they are executed one by one on the new device, and with their performance results collected. After this, a new column is added to the matrix $E$ associated to the new device, with the entries corresponding to each benchmarked technique filled. Next, the new matrix is normalized, followed by applying the previously explained SVD algorithm \ref{subsec:1-1}, where eq.\ref{eq:svd}, \ref{eq:p}, \ref{eq:q}, and \ref{eq:epsilon} are used for error minimization in eq. \ref{eq:min}. The missing performance estimation is then obtained, out of all the registered techniques in the system, saved in the \textit{ML techniques store}, which allows us to scale up, migrate or replicate using the new hardware device.

%\hl{In previosu subsection we say: The utilization of the previously described algorithm to estimate the performance of a ML-based technique is described in the next subsection. But this is not the case here. This subsection woudl benefit from some cross references with the previsous section like pointing to a formula or a statement or else they read rather disconnected.}

\section{Performance Evaluation} \label{sec:evaluation}
To evaluate the performance of our proposed ML algorithm, an online store of AI benchmark experiments is used \cite{aibenchmark2019}. % The database was first released in 2018 to measure the AI performance of various mobile devices. 
It contains numerous AI tasks and deep learning architectures, tested on multiple hardware platforms. Most popular DL architecture like \textit{MobileNet-V2} (classification), \textit{Inception-V3} (classification), \textit{VGG-19} (image-to-image mapping) and \textit{LSTM} (sentence sentiment analysis), and \textit{DeepLab} (image segmentation) are evaluated in terms of training time and inference time (per one image), as well as memory utilization. An updated dataset from 2019 containing 42 different AI tasks built for desktops is used in our work \cite{ignatov2019ai}. Popular hardware platforms, such as CPUs, GPUs and TPUs, are configured to run deep learning models.  In the store, 192 different hardware platforms are tested, e.g.  \textit{Tesla V100 SXM2 32Gb}, \textit{NVIDIA TITAN V},\textit{ GeForce GTX 1080 Ti}, \textit{AMD Threadripper 3970X}, \textit{Intel Xeon Gold 6130}, etc.

To emulate the behavior of the proposed ML recommender, we assume that our system consists of 191 heterogeneous hardware units, and an \textit{ML techniques store} of 42 AI tasks. A new hardware is added to our network, which needs to be tested in order to collect the performance of all the existing AI tasks in the store. The \textit{knowledge base} of the intent-based orchestrator is initialized with a subset of the existing AI tasks and a subset of the existing hardware platforms.  The \textit{ML Recommender} is triggered throught the \textit{dashboard} using the following intents:
 \textbf{"add ML-technique $\textit{MobileNet-V2}$ to ML-technique type  $\textit{threat-detection}$"}, where the \textit{ML-tech-id} is replaced by a different ML technique from the store for every replication e.g. \textbf{\textit{Inception-V3}, \textit{VGG-19}}, etc. After that SVD algorithm runs to predict the benchmark values of missing hardware devices. The metric used to evaluate the machine learning algorithm is the normalized root mean squared error (RMSE) and can be expressed as follows:
\begin{equation}
\text{RMSE} = \sqrt{\frac{\sum_{i=1}^{N}\sum_{j=1}^{N}\epsilon_{i,j}^2}{N^2}}
\end{equation}

\begin{equation}
\text{normalized RMSE} = \frac{\text{RMSE}}{\max{R}-\min{R}}
\end{equation}
where R is the matrix that contains predictions, $\max{R}$ and $\min{R}$ represent the maximum and minimum values in the matrix.

\begin{figure}
 \centering
   \includegraphics[scale=0.4]{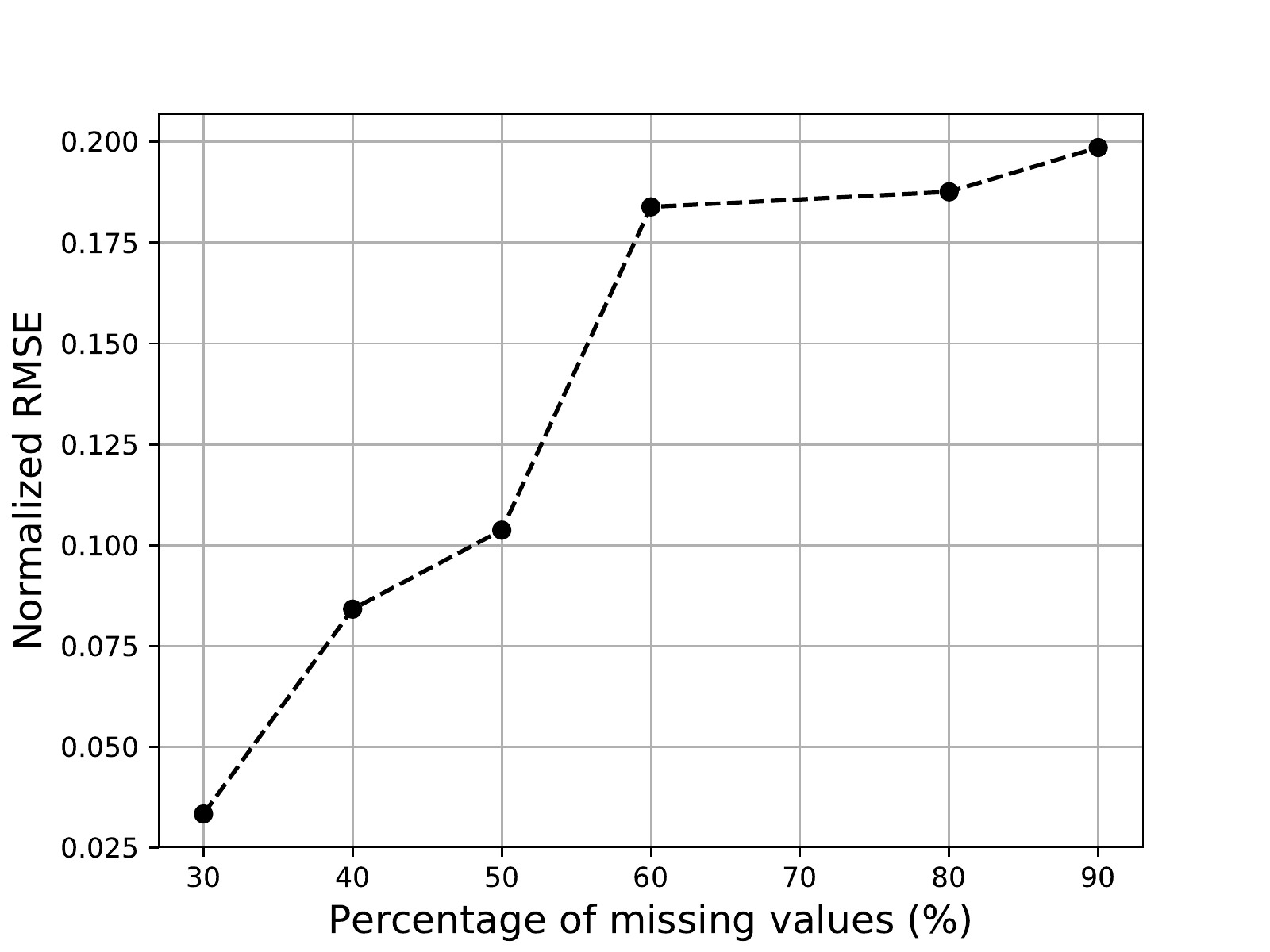}
 \caption{Performance of benchmark prediction using Normalized RMSE with different settings of missing benchmarks.}
\label{fig:missingvalues}
\vspace{-0.2 cm}
\end{figure}

The number of iterations is set to 5000, the latent factors are set to 10, learning rate is 0.04, and  $\beta=5 \cdot 10^{-6}$. In figure \ref{fig:missingvalues} the percentage of missing benchmark is varied from 30\% to 90\%. For each test, 5 replications are evaluated where a random device is chosen for benchmark prediction for every replication. Then the average normalized RMSE is calculated and plotted. The performance of prediction is expected to decrease (normalized RMSE increase) when the percentage of missing values increases. Thus, the more benchmarks we obtain from real measurements, the better our estimation of the AI tasks performance on other hardware devices. For a 30\% of missing devices, the normalized RMSE is equal to 0.03, which is a promising results. However the normalized RMSE for 90\% of missing benchmarks is very high around 0.2, which needs to be improved, in order to use our system for practical applications, where benchmarking techniques are time consuming, and when scheduling must happen in real time using accurate performance evaluations.

%\hl{the perf section is a little too short. We need to expland on implementation and intents implemented if we cannot show anything else. Some discussions would also be good to link the verification of the algorithm itself to the quality fo the solution. 
%the curve can be smoothenend or approximated. How do we scale the reivewer will ask for the restuls? this needs to be commented on. Again, this section needs at least one more para and perhpas a few back of the envelope calculations. Otheriwse the paper reads like much talk about little outcome.}
\section{Conclusion and Future work} \label{sec:conclusion}

With the rise of heterogeneous devices in complex ICT systems and the rise of machine learning based solutions that are implemented to solve problems in every domain of application, it is becoming an imperative to study the challenges of managing ML solutions running in such systems in an innovative way. We first proposed intent-based networking (IBN) solution as the approach for intelligent managing of ICT supply chain systems. Afterwards, we studied the problem of benchmarking these ML solutions, through a  module called \emph{ML Recommender}. This module is used for estimating  the computational performance  of ML-based techniques on heterogeneous hardware devices, considering the lack of explicit knowledge about the number of operations, framework, or the device processing characteristics for each ML-based technique. Collaborative Filtering  techniques like Singular Value Decomposition (SVD) are used to estimate the missing values of the benchmark input  matrices, which is planned to be used to enhance the management system capabilities of scheduling, planning or hardware provisioning. A  dataset of 42 ML model benchmarked on 196 different hardware platform were used for testing the proposed algorithm, where the results are  promising  in terms of estimation correctness. As a future work, we plan to extend our system design of the ML recommender to consider the scheduling and planning problem, and to show how can a user requirement given as a high-level intent be automatically translated into scheduling decisions using our benchmarking solution.

%However, the idea of how to specifically use different machine learning in F2C architecture is still new and evolving.

%(Future work : model a F2C architecture mathematically considering several parameters like processing capabilities of each node in different layers, link capacity, latency,... and compare an ML based strategy with a non ML based strategy  to optimize bandwidth and decrease packet loss))

\section*{Acknowledgment}
This work is  partially funded by European Commission under the H2020-952644 contract for project FISHY: A coordinated framework for cyber resilient supply chain systems over complex ICT infrastructures.

\bibliographystyle{IEEEtran}
\bibliography{mybib}

%\appendix

\end{document}